\documentclass[aps,prl,reprint,groupedaddress,amssymb,twocolumn,superscriptaddress,letterpaper]{revtex4-2}

\usepackage{graphicx}
\usepackage{amsmath}
\usepackage{multirow} 
\usepackage{natbib}
\usepackage{hyperref}
\usepackage{xcolor}

\begin{document}


\newcommand{\muskan}[1] {{\textcolor{cyan}{{#1}}}}
\title{The Very Late Time Afterglow of GW170817 Favors a Wobbling Jet}


\author{Hao Wang}
\email[]{haowang@pmo.ac.cn}
\affiliation{Key Laboratory of Dark Matter and Space Astronomy, Purple Mountain Observatory, \\
Chinese Academy of Sciences, Nanjing 210023, China}

\author{Ore Gottlieb}
\affiliation{Department of Physics and Kavli Institute for Astrophysics and Space Research, Massachusetts Institute of Technology, Cambridge, MA 02139, USA}

\author{Aman Katira}
\affiliation{Indian Institute Of Technology Kanpur, Kanpur, Uttar Pradesh 208016, India}

\author{Muskan Yadav}
\affiliation{Dipartimento di Fisica, Università di Roma Tor Vergata, 
Via della Ricerca Scientifica 1, 00133 Rome, Italy}

\author{Lei Lei}
\affiliation{Key Laboratory of Dark Matter and Space Astronomy, Purple Mountain Observatory, \\
Chinese Academy of Sciences, Nanjing 210023, China}
\affiliation{School of Astronomy and Space Science, University of Science and Technology of China, \\
Hefei 230026, China}

\author{Yi-Zhong Fan}
\affiliation{Key Laboratory of Dark Matter and Space Astronomy, Purple Mountain Observatory, \\
Chinese Academy of Sciences, Nanjing 210023, China}
\affiliation{School of Astronomy and Space Science, University of Science and Technology of China, \\
Hefei 230026, China}

\author{Da-Ming Wei}
\affiliation{Key Laboratory of Dark Matter and Space Astronomy, Purple Mountain Observatory, \\
Chinese Academy of Sciences, Nanjing 210023, China}
\affiliation{School of Astronomy and Space Science, University of Science and Technology of China, \\
Hefei 230026, China}



\begin{abstract}
GW170817 remains the only binary neutron star merger detected through multimessenger emission. Its afterglow has been monitored for nearly a decade, offering an unprecedented opportunity to probe the properties of the outflow. The shallow decay of the very late-time afterglow challenges the prediction of a collimated structured jet. Motivated by recent general-relativistic magnetohydrodynamic simulations, we propose that the GW170817 afterglow is powered by a wobbling jet that drags a ring on the sky. This structure predicts a post-break decay rate shallower than that of a collimated jet, as observers will see a progressively longer emitting arc after the break. A misaligned ring-shaped jet can therefore self-consistently explain the multimessenger data without invoking any extra component. 
Through a Bayesian analysis of the multimessenger data, we find a ring-shaped jet is favored over a collimated jet at a significance level of 4.8$\sigma$. Our results imply a wobbling angle of $\sim 27^\circ$. Such a large angle points to a significant disk tilt, potentially arising from disk-infalling gas interaction or asymmetric angular momentum ejection. Similar shallow decays have also been found in other GRB afterglows, raising the possibility that wobbling jets are common among GRBs.
\end{abstract}


\maketitle

{\it Introduction.}---GW170817 is the only binary neutron star (BNS) merger event in which gravitational waves, Gamma-ray burst (i.e., GRB 170817A), and kilonova have been simultaneously detected \cite{2017PhRvL.119p1101A,2017ApJ...848L..12A,2017ApJ...848L..13A}.  GRB 170817A 
is also the first one evidently observed from a misaligned direction \cite{Troja2017,2018MNRAS.479..588G,2018PhRvL.120x1103L,2021ARA&A..59..155M}. In this scenario, the jet structure becomes important, because its afterglow is initially dominated by the jet ``wing'' component outside the core \citep{2018ApJ...857..128J,2017MNRAS.472.4953L}. A collimated structured jet has successfully explained the rising phase of the afterglow as well as the apparent superluminal motion measured by Very Long Baseline Interferometry \cite{2018Natur.561..355M,2019Sci...363..968G,2022Natur.610..273M}. It also provides an independent constraint on the jet inclination angle, which helps break down the degeneracies in the gravitational wave parameter estimation, and enhances the standard siren measurement of the Hubble constant \cite{2017Natur.551...85A,2019NatAs...3..940H,2021ApJ...908..200W}.

This remarkable event has been continuously monitored in multiple wavebands for nearly a decade\citep{Hallinan2017,Troja2017,Arcavi2017,Alexander2018,2018MNRAS.478L..18T,Troja2019,Troja2020,Balasubramanian2021,Mooley2022,Troja2022,Balasubramanian2022,2025MNRAS.539.2654K}, making it one of the few GRBs with well-measured afterglow emission lasting $\geq 5$ years after the burst, providing an unprecedented opportunity to test jet models. Latest studies have shown that a collimated and structured jet (hereafter a collimated jet) has difficulty explaining the very late time afterglow (e.g., \cite{2024MNRAS.528.2600G,2024ApJS..273...17W,2025MNRAS.539.2654K}). In the standard afterglow model, a jet will be fully visible and start to spread sideways after the jet break \cite{1999ApJ...525..737R}, leading to a temporal decay $F_\nu\propto t^{-p}$, where $p>2$ is the electron power index. A broken power-law fitting to the GW170817 afterglow shows that the temporal decay after the peak follows $t^{-\alpha}$ where $\alpha \lesssim 1.8$ \cite{2025MNRAS.539.2654K}. This decline rate is much shallower than the prediction and is only possible for either a very wide jet or a very shallow wing \cite{2023MNRAS.524L..78G}. On the other hand, a collimated jet with a steep wing is strongly supported by the observation of the apparent superluminal motion \cite{2022Natur.610..273M}. This inconsistency challenges our understanding of the jet structure and the sideways spreading process.

Recent jet-launching general-relativistic magnetohydrodynamic (GRMHD) simulations provide clues to this problem. In both collapsar and neutron star black hole (NSBH) merger simulations \cite{2022ApJ...933L...9G,2023ApJ...954L..21G,2023PhRvD.107l3001H}, the accretion disk around the black hole is found to be tilted. In collapsars it originates from the interaction between the infalling gas and the disk, while in binary mergers it comes from asymmetric angular momentum ejection. As a result, the jet associated with a tilted disk wobbles.
In this work, we propose that accretion disks in BNS mergers are also tilted as in collapsars and NSBH mergers, though a self-consistent jet-launching GRMHD simulation in a BNS merger remains a subject for future studies.
Depending on the intermittency of the central engine, a wobbling jet may drag a ``band" on the sky or break into many jet elements. This complicated outflow structure is expected to produce distinct afterglow signatures compared to a collimated jet, since many outflow components are inevitably misaligned. Likewise, \citealt{2025ApJ...992L...3G} recently proposed that the multiple-jet scenario may be responsible for the rebrightening seen in the afterglow of Einstein Probe (EP) transients \cite{2025ApJ...992L...3G}.

In this letter, we propose that the GW170817 afterglow is produced by the misaligned observation of a wobbling jet that drags a continuous band in the sky. To build a toy model, we approximate it using a ring-shaped jet structure. This structure leads to a distinct post-jet break decay since observers will see a progressively longer ``arc" of the jet as the beaming angle increases, resulting in a shallow temporal decay. We find that the analytic decaying rate agrees well with the observations. Meanwhile, the superluminal motion is not affected as long as the ``wing" structure surrounding the ring remains similar to that of a collimated jet. Therefore, this model provides a compelling explanation for the multimessenger data as it does not invoke any additional components or physical processes beyond the standard GRB afterglow. 

{\it The afterglow of a wobbling jet.}---Jet precession is widespread across many astrophysical systems in the Universe, such as supermassive black holes \cite{2023Natur.621..711C} and tidal disruption events. It has long been proposed that GRB jets have precession as well\cite{1999ApJ...520..666P,2006A&A...454...11R,2010A&A...516A..16L,2013PhRvD..87h4053S}. Indeed, first-principle collapsar and NSBH merger simulations have found that GRB jets wobble during their launch\cite{2022ApJ...933L...9G,2023ApJ...954L..21G,2023PhRvD.107l3001H}. Unlike other jet systems, these simulations show that the wobbling of the GRB jet is mainly caused by the disk tilt due to the interaction between the accretion disk and the infalling gas.

The wobbling jet may yield different observational signatures, depending on the relation between the wobbling angle $\theta_{\rm wo}$, the jet half-opening angle $\theta_c$, and the intermittency of the central engine. i) If the wobbling angle is smaller than the jet half-opening angle, $\theta_{\rm wo}<\theta_c$, the wobbling motion is not important, and the jet structure resembles a collimated jet. ii) If the wobbling angle is larger than the opening angle but the central engine is highly intermittent, the jet will break into many jet elements launched in different directions (see \cite{2025ApJ...992L...3G}). The jet structure can then be approximated by multiple independent jets. iii) If the central engine activity is continuous relative to the baryon mixing timescale, the jet will drag a band in the sphere. 

\begin{figure}[ht!]
\includegraphics[width=\columnwidth]{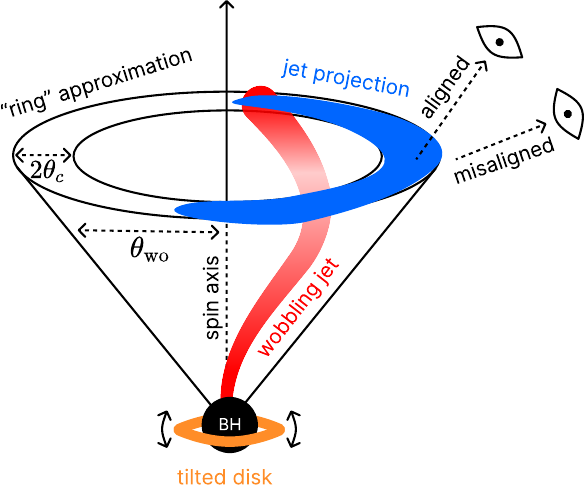}
\caption{The cartoon illustrates our wobbling jet model and its ring-shape approximation.\label{fig:ring}}
\end{figure}

Figure~\ref{fig:ring} illustrates the last scenario. The typical size of the band depends on the wobbling angle $\theta_{\rm wo}$, and the thickness of the band depends on the half-opening angle $\theta_c$ of the initially launched jet. An observer is considered aligned or misaligned depending on whether the line of sight points to the band. In practice, the jet wobbling is not necessarily periodic, symmetric, or closed, and the projected band on the sphere is not necessarily a circle. However, a ring-shaped jet could serve as a toy model, since an asymmetric jet is beyond the capability of all current afterglow modeling tools. Later on, we will show that this toy model is, in practice, a very good approximation.

The temporal behavior of the afterglow for a ring-shaped jet can be derived in the following way. For simplicity, we assume a constant circumburst density $n_0$ throughout this work. The dynamics of the blastwave can be derived from energy conservation. The beaming-visible energy for the blastwave with Lorentz factor $\Gamma$ at radius $R$ is 
\begin{equation}\label{eq:dynamics}
    \frac{1}{3} R^3 n m_p c^2\Gamma^2\Omega =fE_{\rm jet},
\end{equation}
where $\Omega$ is the solid angle of the visible part of the blastwave, $E_{\rm jet}$ is the beaming-corrected energy of the jet, and $f$ is the fraction of energy visible within the solid angle $\Omega$. For ultra-relativistic explosions, the radius of the blastwave can be approximated by $R\sim ct$. The observer's time can be converted to the lab frame time using the equal-arrival-time surface. For aligned observers, this conversion can be approximated as
\begin{equation}\label{eq:eats}
    t_{\rm obs}\simeq \frac{R}{2\Gamma^2}.
\end{equation}
The observed luminosity can be computed by integrating the emissivity over the visible emitting region:
\begin{equation}\label{eq:luminosity}
    L_{\nu} =\int\epsilon'_{\nu'}\delta^3 \Delta_{\rm FS}' R^2 d\Omega \simeq \epsilon_\nu R^3\Omega,
\end{equation}
where $\epsilon_\nu$ is the emissivity, $\delta$ is the Doppler factor, $\Delta_{\rm FS}\approx R/\Gamma^2$ is the thickness of the post-shock region, and the primed variables are evaluated in the fluid comoving frame. The emissivity is usually approximated by a broken power-law with respect to some break frequencies. In the next derivations, we assume a frequency band $\nu_m<\nu<\nu_c$, where $\nu_m$ is the frequency corresponding to the electron minimum energy and $\nu_c$ is the synchrotron cooling frequency. This is also the observed frequency band for the GW170817 afterglow throughout the entire observation. In this case, the spectrum is $\epsilon_{\nu}=\epsilon_{\rm \nu,max}(\nu/\nu_m)^{-(p-1)/2}$ in the observer's frame. A detailed derivation for the full frequency band is presented in the Appendix. Based on the standard afterglow theory, one has $\epsilon_{\rm \nu,max}\propto\Gamma^4$ and $\nu_m\propto \Gamma^4$. Now, the temporal behavior of the afterglow is completely determined by the above equations. 

The evolution of the ring afterglow consists of three stages. Initially, the beaming angle $\theta_b\equiv1/\Gamma$ is smaller than the half-opening angle $\theta_c$, so we have $\Omega\approx\pi\theta_b^2\propto\Gamma^{-2}$ and $f\propto\theta_b^2\propto\Gamma^{-2}$. 
In that case, the afterglow follows the standard deceleration phase of a collimated jet.

After the beaming angle exceeds the half-opening angle, two effects become important. First, the full width of the ring becomes visible, and observers will see a progressively longer arc. In this stage, the fraction of visible energy in the beaming cone is proportional to the arc length: $f\propto \theta_b\propto\Gamma^{-1}$. Second, the ring starts to spread sideways. Assuming sound speed spreading, the thickness of the ring will increase following $\theta_c\approx c'_s/\Gamma\propto\Gamma^{-1}$ in the observer's frame. The sideways spreading along the ring is not important since the fluid elements have so far been causally connected only perpendicular to the ring. As such, the solid angle of the blastwave within the beaming cone is $\Omega\approx\pi\theta_c\theta_b\propto\Gamma^{-2}$. Now, from Eqs.\ref{eq:dynamics} and \ref{eq:eats}, one can derive the scaling relations $R\propto \Gamma^{-1/3}$ and $\Gamma\propto t_{\rm obs}^{-3/7}$. The temporal behavior in this stage can then be derived from Eq. \ref{eq:luminosity}, which is $L_\nu\propto t_{\rm obs}^{-3(2p-1)/7}$. 

Finally, the full ring becomes visible, and the ring merges into a solid wide jet. Now We have $\Omega\propto\pi\theta_c^2\propto\Gamma^{-2}$ and $f\propto 1$. The afterglow is nothing different from the post-jet break stage for a collimated jet.

Although the first and third stages align with the predictions for a collimated jet, a ring-shaped jet has a new intermediate stage. In this stage, the decay index is $3(2p-1)/7$, which is numerically 1.3 for $p=2$ and 1.7 for $p=2.5$. From the above derivation, it is easy to realize that this unique decaying rate is the consequence of the scaling relation $f\propto\theta_b$. This relation indicates that, while the visible region in the beaming cone increases in two dimensions, the amount of energy entering this cone only increases in one dimension. This condition is satisfied as long as the visible emitting region is long and thin, such that one of its dimensions is fully visible. Therefore, the jet structure is not necessarily a closed symmetric ring but a thin band of any shape. 
The afterglow of this band can thus be approximated by that of a symmetric ring whose length and thickness are comparable to the band. As such, a ring-shaped jet is practically a very good approximation. It should also be noted that if the jet wobbling angle is small, or if it is observed from a misaligned direction, the decaying slope could be slightly steeper, because the arc in the beaming cone is bent and its length could be smaller than $\theta_b$. 

The temporal scaling for the afterglow of a ring-shaped jet has previously been studied in the literature \cite{2005ApJ...631.1022G,2023ApJ...957...29D}. However, previous works mostly focus on the hollow-core jet structure (i.e., a thick ring) found in numerical simulations assuming axisymmetry, or equatorial explosions in some supernovae \cite{2022ApJ...931L..16D}. In this work, we propose that the ring structure is caused by the wobbling or precession of a jet. This situation is distinct from previous cases because the ring can be arbitrarily thin and located at an arbitrary latitude. The ring is not necessarily symmetric or closed. 

{\it The GW170817 multimessenger study.}---We now proceed to fit the multimessenger data of GW170817 by our ring-shaped jet model. As mentioned before, the shallow decay of the GW170817 afterglow becomes a problem when fitted together with the apparent superluminal motion, where the latter observation indicates a highly collimated jet. This inconsistency not only challenges our understanding of the jet structure, but also affects the constraint on the observing angle \cite{2024MNRAS.528.2600G}. The observing angle, on the other hand, can be independently measured by gravitational wave data. For this reason, a completely self-consistent study must simultaneously include the afterglow \cite{2021ApJ...922..154M,2025MNRAS.539.2654K}(see references therein), the apparent superluminal motion\cite{2018Natur.561..355M,2019Sci...363..968G,2022Natur.610..273M}, and the gravitational wave data \cite{2017PhRvL.119p1101A}. In particular, this data set includes recent late-time afterglow observations in the radio bands \cite{Balasubramanian2021,Troja2022,Balasubramanian2022} (PIs: Balasubramanian, Margutti, Corsi, Troja) and X-ray bands \cite{Troja2020,2025MNRAS.539.2654K} (PIs: Margutti, Troja), which help to better constrain the decay rate. In this study, the constraint from the gravitational wave data is included by using the publicly available posterior samples\footnote{\url{https://gwosc.org/eventapi/html/GWTC-1-confident/GW170817/v3/}} of the inclination angle and the luminosity distance. We assume that the BNS orbital axis aligns with the wobbling axis, so that the inclination angle is the same as the observing angle. Therefore, the posterior distribution of gravitational wave data can be applied directly as a prior distribution in our multimessenger study.

We perform a Bayesian analysis of the full aforementioned data set. The details of the analysis are discussed in the Appendix. In this study, we consider two models: a jet with a power-law wing structure and a ring-shaped jet. The angular structure for a ring-shaped jet can be easily generalized from a power-law jet \citep{2021MNRAS.500.3511G} in the following way
\begin{align}
    & E_{\rm iso}(\theta)=E_0\left[1+\left(\frac{\theta-\theta_{\rm wo}}{\theta_c}\right)^2\right]^{-s/2}, \\
    & \Gamma(\theta)=(\Gamma_0-1)\left[1+\left(\frac{\theta-\theta_{\rm wo}}{\theta_c}\right)^2\right]^{-s/2} + 1.
\end{align}
This structure degenerates into a power-law jet when the wobbling angle $\theta_{\rm wo}$ is 0. We have also included the Deep Newtonian correction\cite{2003MNRAS.341..263H,2013ApJ...778..107S} in both models, as suggested by previous studies\cite{2024ApJ...975..131R} (see Appendix). 

The dynamics of the blastwave is modeled by the recently developed code \texttt{jetsimpy} \cite{2024ApJS..273...17W,2025ApJ...990..110W}. This code evolves the blastwave using hydrodynamic simulations, which are more accurate than semi-analytic models that assume sound speed spreading \cite{2024MNRAS.531.1704G}. The ring structure has a large hollow core, so the jet will spread both outward and inward. This process is non-trivial for a semi-analytic model to describe because the fluid elements have different spreading directions at different latitudes. A semi-analytic model also has difficulty in describing the interaction of fluid elements when they collide at the pole.

\begin{figure}[ht!]
\includegraphics[width=\columnwidth]{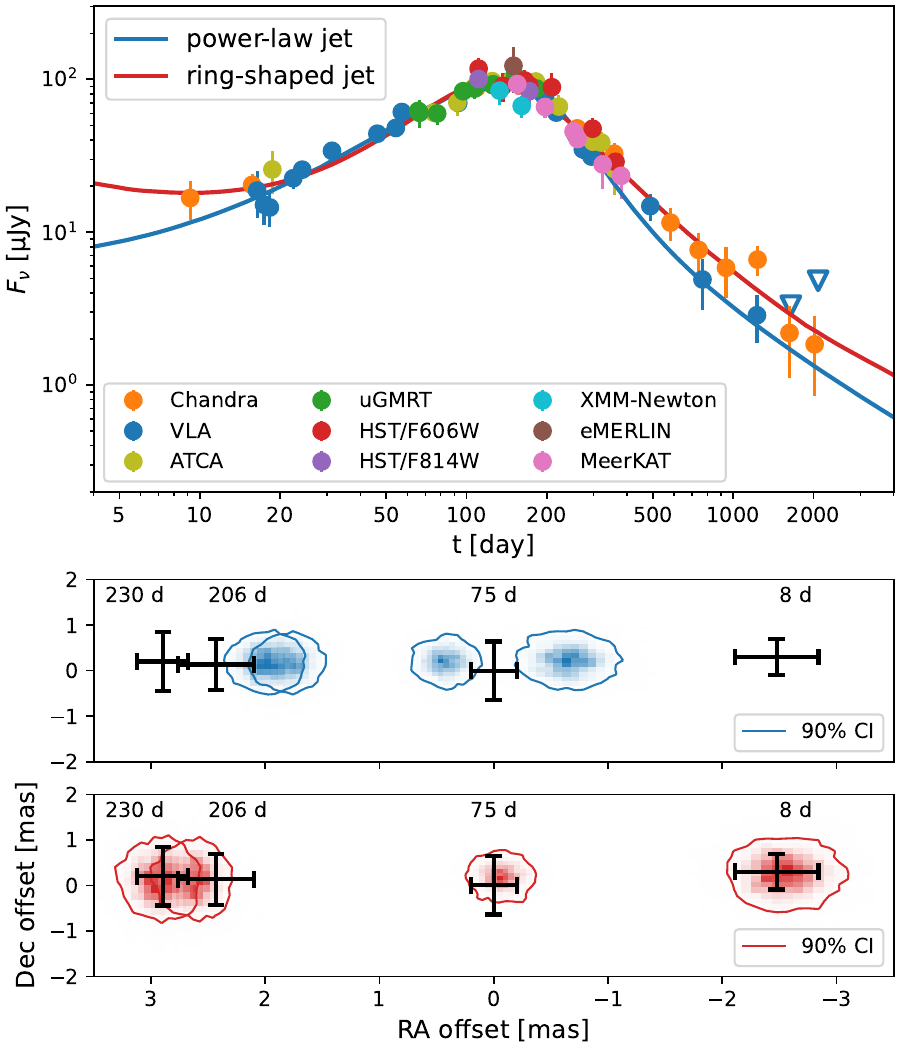}
\caption{Top: the best-fitting light curves for the GW170817 afterglow, where all data are shifted to 3 GHz. Bottom: the 90\% confidence interval for the centroid offset at respective dates. \label{fig:fitting}}
\end{figure}

Figure~\ref{fig:fitting} presents the best fitting results, with the detailed parameter estimation uncertainties listed in Table \ref{tab:parameters} in the Appendix. It clearly shows that a structured jet cannot adequately explain the full data set, consistent with previous findings\cite{2022Natur.610..273M,2024MNRAS.528.2600G}. To compare the goodness-of-fit of the two models, we perform a Bayesian model comparison. We find that a ring-shaped jet is favored over a power-law jet by a logarithm Bayes factor of $\log_{10}\text{BF}=6.16$, corresponding to $4.8\sigma$. Therefore, our result strongly supports a wobbling jet over a collimated solid jet.

{\it Conclusion and discussions.}---In this study, we propose a wobbling jet model to resolve the inconsistency between the GW170817 late-time afterglow and the apparent superluminal observation. Our model is motivated by first-principles GRMHD simulations that show the jet changes its direction during the launching process. We approximate the wobbling jet by a ring-shaped structure as a toy model. We show analytically that the temporal decaying rate after the jet break for a wobbling jet is $F_\nu\propto t^{-3(2p-1)/7}$, shallower than $t^{-p}$ for an axisymmetric collimated jet. The misaligned observation of a ring-shaped jet can therefore simultaneously explain the afterglow, the apparent superluminal motion, and the gravitational wave data. Through a Bayesian data analysis, our model is favored over a structured jet by a logarithmic Bayes factor of $\log_{10}\text{BF}=6.16$, corresponding to $4.8\sigma$. Our findings suggest that GRB jets are wobbling or precessing as in other astrophysical systems. The inferred wobbling angle in our study is $\sim 27^\circ$, which is relatively extreme for BNS mergers. This wobbling motion may indicate significant disk tilt, which may originate from disk-infalling gas interaction or asymmetric angular momentum ejection.

Except for our model, several other models have been proposed in the literature to resolve the afterglow inconsistency, including the possible emergence of the Kilonova afterglow \cite{2023ApJ...943...13W}, refreshed shocks by a stratified radial structure \cite{2025ApJ...987..178S}, a less significant sideway spreading process\cite{2025MNRAS.539.2654K}, or time-dependent microphysical variables\cite{2025MNRAS.539.2654K}. However, the Kilonova afterglow and the refreshed shock models invoke additional components that are not evident in the observation. 
The weak spreading effect is less motivated by hydrodynamics, and time-dependent microphysical variables require extreme parameter evolutions.
Compared to these models, our model is directly motivated by GRMHD simulations and has only one additional parameter, $\theta_{\rm wo}$. In the spirit of Occam’s razor, we therefore conclude that a wobbling jet, as a minimal extension to the structured jet model, is the most natural explanation for the GW170817 multimessenger data.

The wobbling jet model has important implications for a wide range of fields. First, this structure may help resolve the ``missing jet-break" problem. \citealt{2023SciA....9I1405O} has listed a number of GRB afterglows without prominent jet breaks whose decaying slope is around 1.4-1.8. This rate is interestingly consistent with the prediction from a ring-shaped jet. Therefore, wobbling jets may commonly exist among GRB events\cite{2022ApJ...933L...9G}, including the brightest of all time GRB 221009A (Wang et al. in preparation). 

Second, our inferred observing angle for the GW170817 afterglow is larger compared to a collimatted jet model. This leads to a reduced distance constraint $d_L=36.4^{+3}_{-4}$ Mpc, which affects the standard siren measurement of the Hubble constant. Assuming the redshift is $z=0.0098\pm 0.00079$\cite{2017ApJ...848L..31H}, the Hubble constant is $H_0=80.8_{-9.2}^{+11.3}$ km/s/Mpc. This value is not yet accurate enough to resolve the Hubble tension, but slightly favors the value inferred by Type 1a supernova\cite{2022ApJ...934L...7R}. We stress that this value may suffer a large systematic error if the wobbling axis is misaligned with the BNS merger orbital axis or if the ring is asymmetric.

Third, a ring-shaped structure also significantly changes the BNS merger rate inferred from short GRBs. In fact, the merger rate inferred from gravitational wave data is found to be lower than that from short GRBs \cite{2022LRR....25....1M,2026arXiv260406772J}. Although the two values remain consistent owing to the large uncertainty, their central values differ by an order of magnitude. If the jet structure is a ring, its solid angle should be $4\pi\theta_c\theta_{\rm wo}$, which could be much higher than $\pi\theta_c^2$ for a collimated jet. Therefore, the BNS merger rate inferred from short GRBs may be overestimated, and a wobbling jet model can resolve this inconsistency. 

Finally, we emphasize that a ring-shaped jet is just a toy model approximation to a wobbling jet. The practical outflow structure revealed by simulations is highly asymmetric and complex\cite{2022ApJ...933L...9G}. We encourage the community to further investigate its afterglow signatures, which requires an efficient three-dimensional afterglow modeling tool.

{\it Data availability}---The posterior samples of our fitting parameters and their corner plots can be found on the GitHub page\footnote{\url{https://github.com/haowang-astro/GW170817-wobbling-jet}}.

\acknowledgments

{\it Acknowledgments}---This work is supported in part by
    the National Natural Science Foundation of China (No. 12588101, No. 12233011, and No. 12473049), and the National Key R\&D Program of China (No. 2024YFA1611704 and No. 2024YFA1611700).

\bibliographystyle{apsrev4-2}
\bibliography{apssamp}

\clearpage

\appendix
\section{Appendix}\label{sec:appendix}
{\it Temporal behaviors of the ring afterglow}---Here we briefly derive the afterglow temporal behaviors for a ring-shaped jet in all wavebands, which might be useful in future works. As mentioned in the main text, the afterglow of a ring-shaped jet is fully determined by eq. \ref{eq:dynamics}, eq. \ref{eq:eats}, and eq. \ref{eq:luminosity}. Here We follow the common assumptions for the synchrotron radiation in the literature, where readers may refer to \citealt{2015PhR...561....1K} for a review. The emissivity depends on the post-shock number density $n'$ and the rest mass excluded post-shock energy density $e'$. These values can be derived from the shock jump condition:
\begin{align}
    & n'=4\Gamma n_0,\\
    & e'=4\Gamma(\Gamma-1)n_0 m_p c^2,
\end{align}
where $n_0$ is the circumburst medium number density. Assuming that a fraction $\epsilon_e$ of internal energy converts to the power-law electron kinetic energy, we can find the electron minimum Lorentz factor
\begin{equation}
    \gamma_m=\frac{p-2}{p-1}\frac{\epsilon_em_p}{m_e}(\Gamma-1).
\end{equation}
Assuming that a fraction $\epsilon_B$ of internal energy converts to the magnetic energy, the magnetic field is
\begin{equation}
    B'=\sqrt{32\pi\epsilon_B\Gamma(\Gamma-1)n_0m_pc^2}.
\end{equation}
The electrons whose speed are faster than Lorentz factor $\gamma_c$ are cooled by synchrotron radiation at a timescale shorter than the acceleration timescale, where the value can be estimated by
\begin{equation}
    \gamma_c=\frac{6\pi m_e\Gamma c}{\sigma_t B'^2 t}.
\end{equation}
The typical frequency emitted by the characteristic electrons are
\begin{equation}
    \nu_i=\frac{3eB'\gamma_i^2}{4\pi m_ec},
\end{equation}
where i=\{m,c\}. Now the emissivity in the comoving frame can be approximated by a broken power-law of the following form.
\begin{align}
	& \epsilon'_{\nu'} = \epsilon'_{\rm \nu,max} \nonumber \\
    & \times
	\begin{cases}
		(\nu' / \nu'_m)^{1/3} & \nu' < \nu'_m < \nu'_c \\
		(\nu' / \nu'_m)^{-(p-1)/2} & \nu'_m < \nu' < \nu'_c \\
		(\nu'_c / \nu'_m)^{-(p-1)/2} (\nu' / \nu'_c)^{-p/2} & \nu'_m < \nu'_c < \nu' \\
		(\nu' / \nu'_c)^{1/3} & \nu' < \nu'_c < \nu'_m \\
		(\nu' / \nu'_c)^{-1/2} & \nu'_c < \nu' < \nu'_m \\
		(\nu'_m / \nu'_c)^{-1/2} (\nu' / \nu'_m)^{-p/2} & \nu'_c < \nu'_m < \nu'
	\end{cases}
\end{align}
The peak emissivity is
\begin{equation}
    \epsilon'_{\rm \nu',max}=\frac{4\sqrt{3}e^3B'\Gamma n_0}{m_e c^2}.
\end{equation}
The above prescription of synchrotron emissivity has an issue when $\epsilon_e$ and $\Gamma$ are very small, which may lead to $\gamma_m<1$. This problem can be resolved by a correction at the deep Newtonian phase\cite{2003MNRAS.341..263H,2013ApJ...778..107S}. When $\gamma_m$ inferred by the above equation is $\gamma_m\lesssim 1$, many electrons are Newtonian whose synchrotron radiation are negligible in the frequency band of interest. In this situation, we fix $\gamma_m=1$ and the fraction of power-law electrons can be estimated by
\begin{equation}
    \xi_n=\frac{p-2}{p-1}\frac{\epsilon_e m_p}{m_e}(\Gamma-1).
\end{equation}
A reduced number of radiating electrons is implemented by adjusting the peak emissivity to $\xi_n\epsilon'_{\rm \nu',max}$. Some studies have suggested that this correction may explain the shallow decay of the GW170817 afterglow \cite{2024ApJ...975..131R}. For this reason, we have included this effect in the data fitting, but we find this correction not important for a ring-shaped jet model. In this analytic study, we will neglect this effect for simplicity.

We can convert the peak emissivity and break frequencies to the observer's frame by $\epsilon_{\rm \nu,max}=\epsilon'_{\rm \nu',max}\delta^2$ and $\nu_i=\nu'_i\delta$, where $\delta$ is the Doppler factor. The spectrum power indices remain unchanged. In the ultra-relativistic limit, the above equations yield $\epsilon_{\rm\nu,max}\propto\Gamma^4$, $\nu_m\propto\Gamma^4$, and $\nu_c\propto R^{-2}$. After the jet break, we have derived in the main text that $R\propto \Gamma^{-1/3}$ and $\Gamma\propto t_{\rm obs}^{-3/7}$. Inserting the above scaling relations to eq. \ref{eq:luminosity}, we can find the temporal behavior in the full waveband. Assuming that $F_\nu\propto t^\alpha$ in respective bands, the result is
\begin{align}
    & \alpha = \begin{cases}
		1/7 & \nu < \nu_m < \nu_c \\
		-3(2p-1)/7 & \nu_m < \nu < \nu_c \\
		-2(3p-1)/7 & \nu_m < \nu_c < \nu \\
		-1/3 & \nu < \nu_c < \nu_m \\
		-4/7 & \nu_c < \nu < \nu_m \\
		-2(3p-1)/7 & \nu_c < \nu_m < \nu
	\end{cases}
\end{align}
The jet break time for a ring-shaped jet happens when the beaming angle exceeds $\theta_c$, which is exactly same as the jet break time for a collimated jet with half-opening angle $\theta_c$. Similarly, the end time of this phase happens when the full ring becomes visible, which is the jet break time for a collimated jet with half-opening angle $\theta_{\rm wo}$. 

{\it Bayesian analysis of the GW170817 afterglow}---Given a model $M$, the Bayesian posterior probability distribution is
\begin{equation}
    p(\vec{\theta};M)\propto \mathcal{L}_{\rm AG}(\vec{d}|\vec{\theta};M)\mathcal{L}_{\rm offset}(\vec{d}|\vec{\theta};M)\pi_{\rm GW}(\vec{\theta}).
\end{equation}
The likelihood of the afterglow $\mathcal{L}_{\rm AG}$ combines the contributions from optical, radio, and X-ray data. The optical and radio likelihood are computed assuming a $\chi^2$ distribution. However, the late time X-ray photon number is very limited, and a Poisson distribution must be applied. The Poisson likelihood of a single X-ray observation is
\begin{equation}
    \mathcal{L}_{\rm x}=\frac{1}{N!}\left[b+a(\vec{\theta};M)\right]^N\exp\left[-b-a(\vec{\theta};M)\right],
\end{equation}
where $N$ is the number of observed photons, $b$ is the expected number of background photons, and $a$ is the computed photon number given an afterglow model $M$. In this study, $a$ is converted from the flux in the following way
\begin{equation}
    a=\mathrm{Flux\ (0.3-10\ keV)}\ /\ \mathrm{ECF}\ \times\ \mathrm{Exposure},
\end{equation}
where ECF is the Energy Conversion Factor inferred from the X-ray data analysis. The total X-ray likelihood is the multiplication of all observations. The parameterization of the apparent superluminal motion follows our previous work, where the trajectory of the centroid is assumed to start from a sky coordinate ($RA_0,Dec_0)$ and moves along an orientation angle of $\theta_{\rm rot}$. Similar to previous studies, the coordinate origin is places at the centroid's position at 75 days. The likelihood of the apparent superluminal motion $\mathcal{L}_{\rm offset}$ is assumed to follow a $\chi^2$ distribution, where both statistic and systematic errors are included. The redshift for this event is assumed to be $z=0.0098$. The prior distributions $\pi(\vec{\theta})$ for the parameters are listed in Table \ref{tab:parameters}. For this specific event, the coasting phase is undetected, so the initial Lorentz factor $\Gamma_0$ cannot be well constrained. For this reason we assume an extremely large value $\Gamma_0\sim10^{10}$ to reduce the parameter dimension.

\begin{table}
	\centering
	\caption{The prior distributions and parameter estimation results for the GW170817 multimessenger study.}
	\label{tab:parameters}
	\begin{tabular}{c|c|c|c} 
		\hline
        \multirow{2}{*}{Parameters}  & \multirow{2}{*}{Prior} & \multicolumn{2}{c}{Parameter Uncertainty} \\
        \cline{3-4}
        & & power-law jet & ring-shaped jet \\
		\hline
		      $\log_{10}(n_0/{\rm cm}^{-3})$ & [-5, 2]& $0.27_{-0.73}^{+0.54}$ & $-3.49_{-0.66}^{+0.64}$\\
            $\log_{10}(E_{\rm iso}/{\rm erg})$ & [49, 57] & $54.29_{-0.72}^{+0.52}$ & $52.88_{-0.67}^{+0.65}$\\
            $s$ & [3, 10] & $8.27_{-1.46}^{+1.15}$ & $3.40_{-0.28}^{+0.51}$\\
            $\theta_{c}$ [rad] & [0, $\pi/2$] & $0.17_{-0.03}^{+0.02}$ & $0.03_{-0.01}^{+0.01}$\\
            $\theta_{\rm wo}$ [rad] & [0, $\pi/2$ & - & $0.47_{-0.09}^{+0.12}$\\
            $\theta_{\rm obs}$ [rad] & GW & $0.60_{-0.04}^{+0.06}$ & $0.73_{-0.11}^{+0.14}$\\
            $d_L$ [Mpc] & GW & $39.44_{-1.91}^{+1.81}$ & $36.43_{-3.76}^{+3.28}$\\
            $\log_{10}\epsilon_{\rm e}$ & [-6, 0] & $-3.20_{-0.55}^{+0.76}$ & $-1.67_{-0.44}^{+0.48}$\\
            $\log_{10}\epsilon_{\rm B}$ & [-6, 0] & $-5.21_{-0.52}^{+0.73}$ & $-3.81_{-1.52}^{+1.61}$\\
            $p$ & [2, 2.5] & $2.12_{-0.01}^{+0.01}$ & $2.16_{-0.01}^{+0.01}$\\
            $RA_0$ [mas] & [-10, 10] & $-0.66_{-0.20}^{+0.21}$ & $-2.55_{-0.25}^{+0.23}$\\
            $Dec_0$ [mas] & [-10, 10] & $0.22_{-0.30}^{+0.29}$ & $0.27_{-0.38}^{+0.37}$\\
            $\theta_{\rm rot}$ [rad] & [-$\pi$, $\pi$] & $-0.01_{-0.13}^{+0.12}$ & $-0.03_{-0.11}^{+0.11}$\\
        \hline
            $\log_{10}\text{BF}$ ($\sigma$) & - & 0.0 (0.0) & 6.16 (4.8) \\
        \hline
	\end{tabular}
\end{table}

In this study we have considered two models: a power-law jet and a ring-shaped jet. The statistical significance of a model is defined as the Bayesian evidence:
\begin{equation}
    Z(M)=\int\mathcal{L}(\vec{d}|\vec{\theta};M)\pi(\vec{\theta})d\vec{\theta}.
\end{equation}
The preference for a model over the other can be described by the Bayes factor, which is the ratio of Bayesian evidences of the two models. It is more convenient to express it by the logarithm Bayes factor, which is $\log_{10}\text{BF}=\Delta\log_{10}Z$. The parameter estimation results and the logarithm Bayes factor are listed in Table \ref{tab:parameters}. 

The posterior sampling and model comparison are performed by a nested sampling algorithm provided by the public package \texttt{Bilby} and \texttt{nessai}. This algorithm is efficient especially when the parameter dimension is high.

\end{document}